\documentclass{article}

\usepackage{fullpage}
\usepackage{graphicx}
\usepackage{latexsym}
\usepackage{amsfonts,amsmath,amssymb}
\usepackage{url}
\usepackage{fancyref}
\usepackage{authblk}
\usepackage{sidecap}
\usepackage[colorlinks=true,pagebackref,linkcolor=magenta,urlcolor=magenta]{hyperref}
\usepackage[sort&compress,colon,square,numbers]{natbib}

\begin{document}

\title{\vspace{-50pt}Stability and localization of inter-individual differences in functional
connectivity}

\author[1]{Raag D.~Airan}
\author[2,3]{Joshua T.~Vogelstein}
\author[1]{Jay J.~Pillai}
\author[4]{Brian Caffo}
\author[5,6]{James J.~Pekar}
\author[1]{Haris I.~Sair}

\affil[1]{Department of Radiology and Radiological Sciences, Johns Hopkins University}
\affil[2]{Department of Statistical Sciences, Duke University}
\affil[3]{Child Mind Institute}
\affil[4]{Department of Biostatistics, Johns Hopkins University}
\affil[5]{F.M. Kirby Research Center for Functional Brain Imaging, Kennedy Krieger Institute}
\affil[6]{Russell H. Morgan Department of Radiology and Radiological Science,
Johns Hopkins School of Medicine}

\maketitle 

%

\vspace{-25pt}

\begin{abstract}
	Much recent attention has been paid to quantifying anatomic and functional neuroimaging on the individual subject level. For optimal individual subject characterization, specific acquisition and analysis features need to be identified that maximize inter-individual variability while concomitantly minimizing intra-subject variability. Here we develop a non-parametric statistical metric that quantifies the degree to which a parameter set allows this individual subject differentiation. We apply this metric to analyzing publicly available test-retest resting-state fMRI (rs-fMRI) data sets. We find that for the question of maximizing individual differentiation, there is a relative tradeoff between increasing sampling through increased sampling frequency or increased acquisition time; that for the sizes of the interrogated data sets, only 4-5 min of acquisition time is necessary to perfectly differentiate each subject; and that brain regions that most contribute to individual’s unique characterization lie in association cortices thought to contribute to higher cognitive function. These findings may guide optimal rs-fMRI experiment design and may aid elucidation of the neural bases for subject-to-subject differences.
\end{abstract}


\section{Introduction}

Many neuroimaging studies seek to use inter-individual variability in anatomic and functional characterization to gain insight into concomitant variability in a behavioral or clinical feature of interest \cite{KanaiR2011,ZillesK2013}. In addition, just as a psychologist may use a patient’s behavioral score to predict an outcome or guide treatment, there has been interest in developing methods and standards that allow for individual patient anatomic and functional characterization using neuroimaging \cite{AtluriG2013}. 

Recent studies have indeed identified significant inter-individual variability in behavioral, anatomic and functional parameters, and determined that these variables correlate in significant and intriguing ways. Kanai and Rees catalogued numerous examples of studies where particular behavioral traits can be predicted by individual subject level region-specific anatomic (DTI, VBM) or functional (BOLD fMRI, PET, MEG, EEG, MRS) measures \cite{KanaiR2011}. Additionally, Mueller et al. demonstrated that there is significant inter-individual variability in functional connectivity assessed with resting state fMRI (rs-fMRI), and that regions of high variability correlate with regions of evolutionarily recent cortical expansion, as well as regions thought to determine higher cognitive function \cite{MuellerS2013}.  

For studies that seek to develop a functional characterization of an individual subject, specific acquisition and analysis features can to be identified that maximize this inter-individual variability while minimizing intra-subject variability. We therefore sought to analyze rs-fMRI data for factors that affect this individual subject differentiation, given the emergence of rs-fMRI as a powerful tool for both neuroscience and clinical application \cite{FornitoA2011, LeeMH2013, SnyderAZ2012}. In particular when compared to task based fMRI, rs-fMRI has received interest as a clinical tool due to (i) potentially low individual variability,  (ii) a lack of dependence on subject compliance for a particular task, and (iii) potentially lower acquisition times \cite{KellyC2012}. The lack of dependence on subject compliance is particularly important in certain patient classes, such as those with functional deficits related to brain lesions, and patients who may not be able to understand the task due to cognitive dysfunction, language barriers, or in the pediatric population.

There are important differences between analysis of task-based and resting-state fMRI data. Without a task with which to correlate, pairwise similarities between regions or voxel timecourses are computed instead of a univariate analysis of the time-series. Additionally, a whole-brain or network based approach is often utilized as there is no prior knowledge to regions of interest. Typically in rs-fMRI connectivity studies, a parameter of similarity (e.g. Pearson correlation, partial correlation, spectral coherence) is calculated between pairs of timecourses from given networks or regions of interest. These similarity data are organized into an adjacency matrix, one calculated per each subject or scan. Fundamentally, each summary statistic, such as ``global efficiency'', calculated in a rs-fMRI study is a calculation of the data in these adjacency matrices \cite{VanDijkKR2010}. 

While group-level reproducibility of rsfMRI is validated for several specific ROIs and networks \cite{BiswalBB2010,ChouYH2012,DamoiseauxJS2006,FrancoAR2013,SongJ2012,WangX2013}, an analysis of whole-brain subject level variability has only recently been described \cite{MuellerS2013}. Here, we focus on acquisition and parcellation factors that most affect the measured inter-individual and intra-individual variation of whole-brain adjacency matrices between test and retest sessions. We develop a non-parametric statistical metric that allows us to quantify the degree to which a given acquisition and analysis scheme maximizes inter-individual variability while minimizing intra-individual variability. We then use this metric to determine which factors appear to most contribute to individual subject differentiation.

\section{Methods}

We would like to utilize a formal notion of reliability that will enable to compute reliability of the data in a justified and principled fashion, and be able to compare the reliability of different procedures.  Figure \ref{fig1} shows a schematic illustration of our method, which we define formally below.
The data that we analyze are subject to many different sources of variability including (i) biological, (ii) scan acquisition, and (iii) graph inference reliability.  We treat each separately, as each is admissible to a different set of properties and manipulations.   Minimizing biological variability is beyond the scope of this work, we therefore focus on understanding and analyzing variability of the scan acquisition and graph inference. That said, because biological variability is clinically useful and important variability, we begin there.  Let $\mathbb{B}(\cdot) : \Omega \to \mathcal{B}$ be a ``brain valued'' random variable, where $\Omega$ indexes both the subject $i$ and the time $t$ of the acquisition. Thus, $B_{i,t} \in \mathcal{B}$ is the brain of subject $i$ at time $t$. For brevity, we define the index $k := k(i,t)$ to indicate brain $i$ at time $t$.

We will study the reliability of these data via analysis of their inferred functional connectomes.  Here, we define a functional connectome to be a weighted graph estimated from fMRI data.  Let $G_v=(V,E) \in \mathcal{G}_v$ be a graph with $|V|=v$ vertices and $E \subset V \times V$ is the collection of edges amongst them.  In an abuse of notation, let $G_v=(V,E,W)$ also denote a weighted graph, where $W=\{w_{ij}\}$ is set of weights.  The weighted adjacency matrix (or adjacency matrix for short), $A \in \mathbb{R}^{v \times v}$ is a $v \times v$ matrix whose elements $a_{ij}$ are $0$ when there is no edge between vertices $i$ and $j$, and $a_{ij}=w_{ij}$ when there is an edge between $i$ and $j$.  Note that these graphs are undirected, so that $a_{ij}=a_{ji}$.  Next we explain how we estimate these functional connectomes.

\subsection{Scan Acquisition Methods}

Formally, let $S_m(\cdot) : \mathcal{B} \to \mathcal{X}$ be the $m^{th}$ acquisition function that takes as input a real brain at some time, and outputs some data $X^m_k = S_m(B_k) \in \mathcal{X}$.  The index $m$ here indexes different scanners and/or acquisition protocols. Note that it cannot be the case that we have both $X^m_k$ and $X^m_{k'}$, because $k$ indexes time as well, so we can drop the superscript for brevity without loss of generality. We consider four different acquisition functions as described as follows:
\begin{enumerate}
	\item \textbf{KKI:} Often referred to as ``KIRBY21'', KKI is a collection of 21 subjects, each scanned twice  at 3T for 7 minutes, TR 2.0 sec \cite{LandmanBA2011}.  One subject was excluded due to data quality.
	\item \textbf{NKI Standard:} A total of 23 subjects acquired at 3T using single band protocols for 5 minutes, TR 2.5 sec \cite{NoonerKB2012}.
	\item \textbf{NKI Multiband TR=1.4 sec:} Same subjects and scanner as above, but using a multiband sequence, and a faster TR of 1.4 sec \cite{NoonerKB2012}.
	\item \textbf{NKI Multiband TR=0.645 sec:} Same as above, but with TR of a faster 0.645 sec \cite{NoonerKB2012}.
\end{enumerate}

Thus, in total, $S_m$ consists of the scanner sequences and pre-processing routines, for $m \in [4]=\{1,2,3,4\}$.  The output of any such $S_m$ is a multivariate time-series, $X_{m,k} \in \mathcal{X} \subset \mathbb{R}^{P \times N_m}$, where $P$ is the number of voxels in the brain, and $N$ is the number of time steps for procedure $m$.   We refer to each such observation, $X_k$, as a ``brain scan''.

\subsection{Graph Inference Methods}

Let $T_p(\cdot) : \mathcal{X} \to \mathcal{G}$ be the $p^{th}$ graph inference routine, that estimates the brain-graph (or connectome) from a brain scan. The graph inference procedures that we consider consist of a sequence of steps.

\paragraph{Shared Pre-processing:}
After the data came off the scanner, we conducted the following pre-processing on each scan independently.  Each BOLD time-series underwent a standard preprocessing sequence of slice timing correction, motion co-registration, spatial normalization to the MNI152 2mm template, detrending via high pass filtering, Principal Component Analysis-based nuisance and motion parameter regression (following \cite{BehzadiY2007}), bandpass temporal filtering (0.1-1.0 Hz) and spatial smoothing (6mm Gaussian kernel). Note that the multiband data sets did not undergo slice timing correction. All analysis was completed in MATLAB (Mathworks, Natick, MA) and SPM8 (Wellcome Trust, UK).

\paragraph{Parcellation Scheme:}
 Two parcellation schemes were tested. For each, the target ROI number for each parcellation scheme was varied in powers of 2 from 128 to 2048. 
Letting $\mathcal{P}$ be the set of voxels, we define a partition by $C$ clusters $\mathcal{P}_1, \ldots, \mathcal{P}_C$, such that $\mathcal{P}_c \cap \mathcal{P}_{c'} = \empty$ for all $c \neq c'$,  $\cup_c \mathcal{P}_c = \mathcal{P}$
For the first parcellation scheme,  the gray matter was subdivided into uniform sized ROIs (maximum size difference between ROIs of one voxel) \cite{ZaleskyA2010}. Thus,  $|\mathcal{P}_c| \approx K_C$ for all $c$, where $| \cdot |$ is the set cardinality operator and $K_C$ is the approximate number of voxels per cell in the partition.
For the second scheme, for the ‘functional’ type, we utilized a recently published scheme based on clustering rs-fMRI data.  The goal of this method was to cluster voxels with highest effective intra-ROI connectivity versus inter-ROI connectivity, while maintaining spatial proximity \cite{CraddockRC2012}. The fMRI data used for generating these parcellations were from three subjects from a different, publicly available data set included in the distribution of the published parcellation code.

\paragraph{Timecourse Extraction Method:}
Regardless of the parcellation scheme used, let $X^c_k=\{X_k(p)\}_{p \in \mathcal{P}_c} \in  \mathbb{R}^{|\mathcal{P}_c| \times N_m}$ be the submatrix extracting just the rows corresponding to voxels in cell $\mathcal{P}_c$. Dropping the subscript $k$ for brevity, we consider two approaches to obtaining univariate time-series from each such matrix.  First, the mean of the ROI time-series, we compute by $x^c_{mean} = \frac{1}{|\mathcal{P}_c|}\sum_{p \in \mathcal{P}_c} X^c$, where $x^c_{mean}=(x^c_{mean} (1), \ldots, x^c_{mean}(N_m)) \in \mathbb{R}^{N_m}$.  Second, let $[D^c,U^c]$ be the eigenvalues and eigenvectors of $X^c$.  Let $x^c_{eig}$ be the projection of $X^c$ onto its first principal component, that is, the eigenvector associated with the largest positive eigenvalue.  Finally, let $Y^{C,\psi}_k \in \mathbb{R}^{C \times N_m}$ be the dimension reduced time-series corresponding to the time-series from observation $k$ with $C$ ROIs using $\psi \in \{mean,eig\}$.

\paragraph{Graph Inference:}
Consider the matrix $Y_k=\{y_{i,t}$, where $y_{i,t}$ is the $i^{th}$ ROI at time $t$. To obtain our connectome estimate, we compute the Pearson correlation matrix:
\begin{align}
	w_{ij} = \frac{ N_m \sum_{t \in [N_m]} y_{i,t} y_{j,t} - N_m \bar{y}_i \bar{y}_j} {(N_m -1) s_i s_j}  
\end{align}
where $\bar{y}_i$ and $s_i$ are the sample mean and standard deviation for ROI $i$.  
Finally, we let $a_{ij}=w_{ij}$ if $w_{ij} > \tau$, and $a_{ij}=0$ otherwise, for some threshold $\tau$.  We only allow $\tau$ to be bigger than $-\infty$ for Figure \ref{fig4} (bottom), where we consider a range of $\tau$'s.  Thus, the graph of observation $k$ will be indicated by the adjacency matrix $A_k=\{a_{ij}^k\}$, and we will use these estimates to conduct our reliability analysis.

\subsection{Reliability Analysis}

\subsubsection{Rank Sum Statistic}

Let $q=(p,m)$ index acquisition and graph inference pair procedure, so that $V_q = T_p \circ S_m$.  The goals of this reliability analysis are to:
\begin{enumerate}
	\item quantify the reliability of various choices procedures $V_q$, and
	\item evaluate whether any $V_q$'s under consideration are sufficient.
\end{enumerate}

We consider a non-parametric, order statistic based strategy.  This strategy has a number of advantages over classical approaches, such as intraclass correlation \cite{ShroutPE2008}.  First, it can operate on multivariate, and even non-Euclidean observations, rather than only real-valued observations.  Second, it is model-free, in that it does not make any assumptions about the distribution of the data.  Third, because it is based on order statistics, it is robust to many kinds of artifacts, such as spurious or missing data.

To proceed, let $\delta_{\cdot,\cdot} : \mathcal{G} \times \mathcal{G} \to \mathcal{R}_+$ be a distance metric between a pair of graphs.  Based on this metric, we defined an order statistic for each observation. 
If the data are reliable, then  $\delta((i,j),(i,t'))$ is relatively small for all $i$.
Let $R^q_k$ be the rank score of observation $k$ for procedure $q$ defined as follows.
For each observation $k$, we sort the observations in decreasing order, such that $\delta_{k,(1)} \leq \delta_{k,(2)} \leq \cdots \leq \delta_{k,(n-1)}$.   Recalling that $k$ indexes a subject and time pair, $R^q_k$ is the rank of observation $i,t'$ with respect to observation $i,t$. 


Ideally, the set of $R^q_k$'s are small for a given procedure. 
We would like to use these order statistics to assess the reliability of different acquisition and graph inference functions.    We define the reliability of $V_q$ as the sum of ranks of all graphs inferred for that procedure. In other words, we let $\mathcal{R}^q = \sum_{k} R^q_k$. Note that $n \leq \mathcal{R}^q \leq n (n-1)$, so this reliability metric is closed and bounded.   

To complete the description of the reliability statistic, we must choose a graph distance metric, $\delta$.  To be in alignment with the neuroimaging community, we choose the Frobenius norm between weighted adjacency matrices:
\begin{align}
	\delta(G_k,G_{k'}) = ||A_k - A_{k'}||_F^2,
\end{align}
where $||X||_F=\sum_{ij} \sqrt{x_{ij}^2}$.

\subsubsection{Permutation Testing}

To assess the reliability of the data, we start by assuming that each graph is sampled independently from some distribution, $A_{i,t} \sim F_{i,t} \in \mathcal{F}$.  If each $F_{i,t}=F$ is the same, then we could not hope to distinguish individual subjects.  However, if the data from each subject were sampled from its own distribution, that is, $A_{i,t} \sim F_i$, assuming that each $F_i$ is sufficiently different from the others, then the data would be reliable. Let $\Delta : \mathcal{F} \times \mathcal{F} \to \mathcal{R}_+$ be a metric computing the distance between densities.  Let $\Delta(F^q_i,F^q_{i'})$ be the distance between distribution $F_i^q$ and $F_{i'}^q$, where $q$ is indexing the procedure for obtaining the graphs, as described above.  The test statistic we consider is therefore $\Delta_q = \max_{i,i'} \Delta(F^q_i,F^q_{i'})$. A reliability analysis can then be thought of in terms of the following hypothesis test:
\begin{align*}
	H_0:& \Delta_q \geq \epsilon \\
	H_A:& \Delta_q < \epsilon.
\end{align*}
Rejecting the null in favor of the alternative implies that we believe that procedure $V_q$ is significantly more reliable than chance.  We can estimate the null distribution via the following permutation test. For $B$ Monte Carlo simulations, we permute the labels, such that $(i,t) \rightarrow (i',t')$.  Because we lack the actual densities $\{F_i\}$, and because estimating high-dimensional densities is a challenging computational statistics problem, we utilize $\mathcal{R}^q$ as a surrogate test statistic. Given the permuted data, it is easy to compute $\mathcal{R}^q_b$ for $b \in [B]$, and therefore obtain an empirical estimate of the distribution of $\mathcal{R}^q$ under the specific null where $\epsilon = 0$.


\section{Results}

\subsection{Stability of Inter-Individual Differences}

\begin{figure}[htbp]
	\centering
		\includegraphics[width=1.0\linewidth]{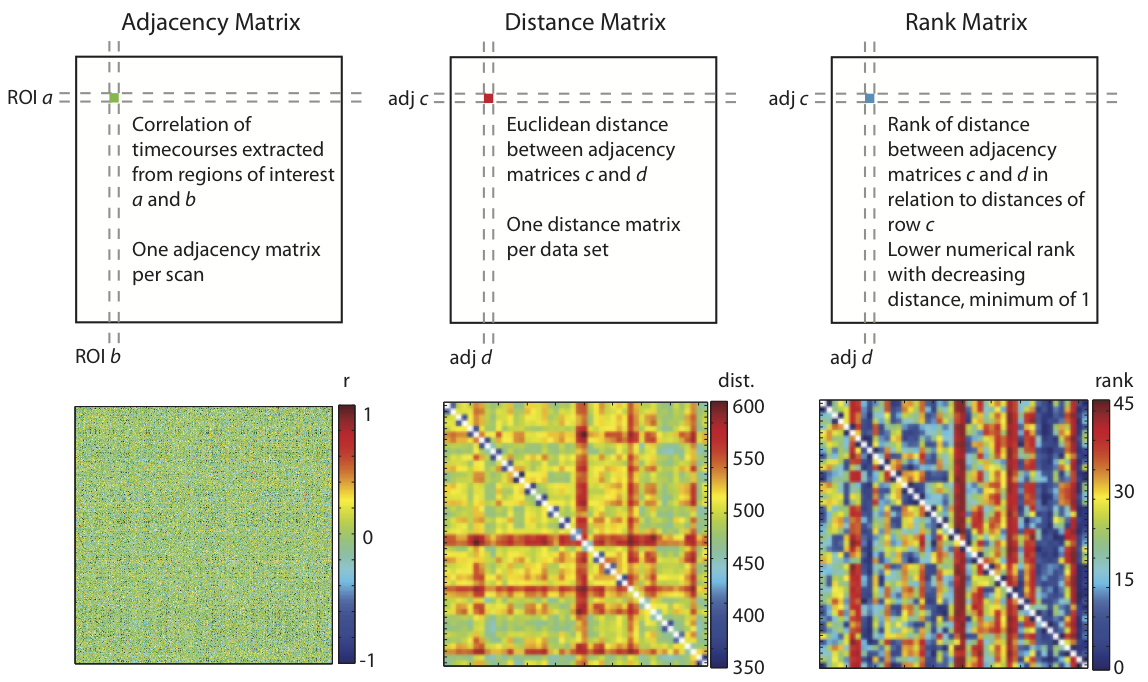}
	\caption{Minimum rank sum statistic enables non-parametric comparisons of test-retest acquisition and analysis parameters.
		To assess analytic efficiency in differentiating individual subjects, first adjacency matrices were made for each scan where each matrix element corresponded to the correlation between time-series extracted from the indicated pair of regions of interest (ROIs; left). Next, for each dataset, a distance matrix was calculated where each element corresponded to the Euclidean distance (square root of sum of square differences for individual matrix elements) between the indicated pair of adjacency matrices (middle). Then, a rank matrix was calculated where each element corresponds to the rank of the distance between the indicated pair of scans compared to the set of distances between that row’s scan and each other scan in the dataset. Note, the rank matrix is not necessarily symmetric (right; minimum rank set to 1).
	}
	\label{fig1}
\end{figure}

A nonparametric rank sum metric was used to compare the ability of different acquisition and parcellation variable values to distinguish individual subjects (Methods; Figure \ref{fig1}). The benefits of using this statistic for comparison include having an absolute minimum (the number of scans in the data set) and easy to define statistics based on random permutation of pairings. This metric is minimized by factors that allow maximal differentiation of individual subjects, with greatest reproducibility between test and retest---in an optimal case the ranks of adjacency matrix distances between test and retest should be 1 for all test-retest pairs, with the rank sum being equal to the number of scans in the data set. 

\begin{figure}[htbp]
	\centering
		\includegraphics[width=0.8\linewidth]{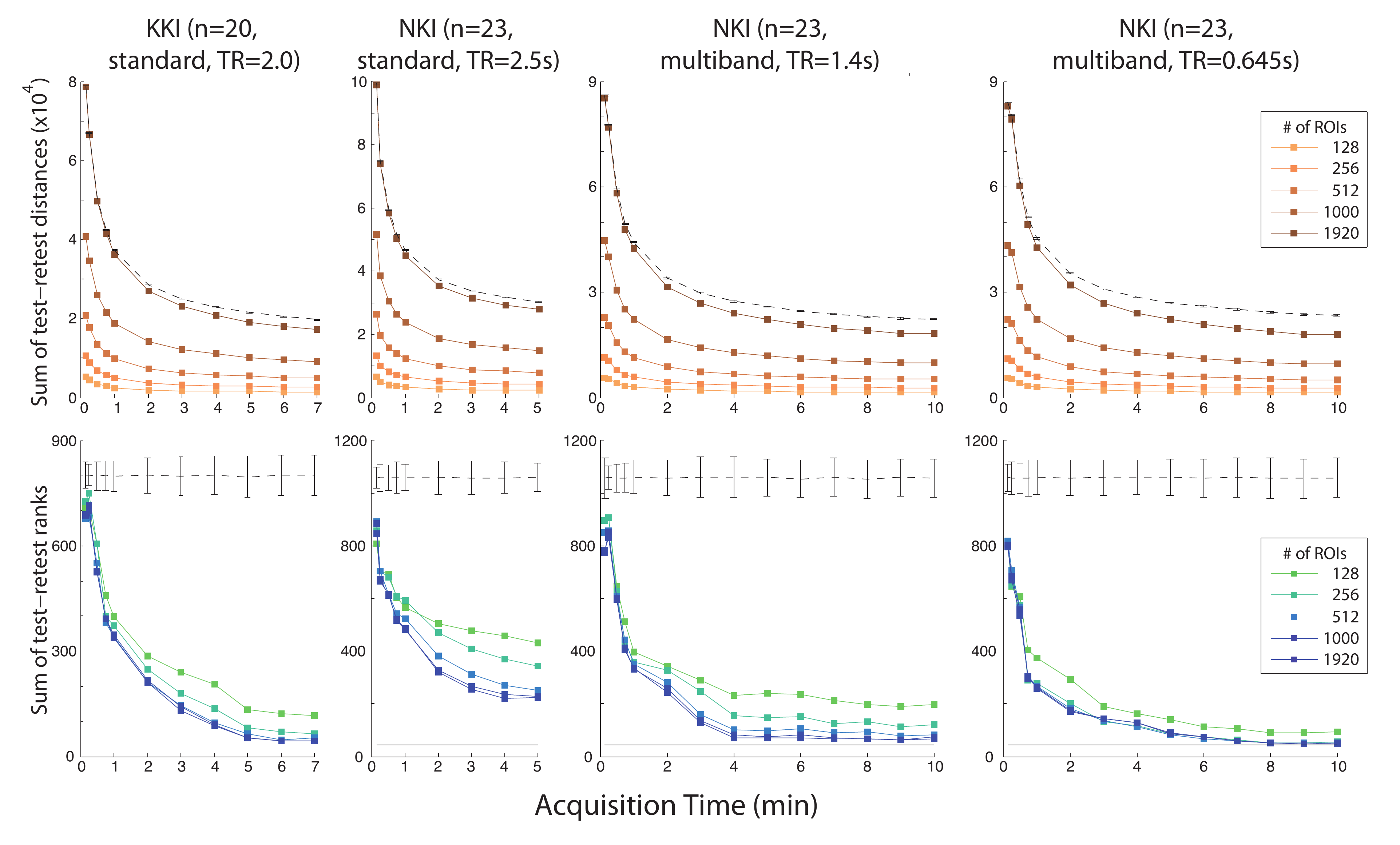}
	\caption{	Varying acquisition time and parcellation for each test-retest data set.
		Top: Sum of distances between test-retest scan pairs versus acquisition time, for varied number of regions of interest in each parcellation (data using a functional parcellation scheme \cite{CraddockRC2012} and eigenvariate time extraction). Dashed line is the mean $\pm$ sem calculation for a set of 1000 randomly assorted pairs for the maximum ROI parcellation.
		Bottom: Sum of ranks of test-retest scan pairs for the indicated data set by ROI, using the same conditions as the Top row. Black line at bottom is the minimum possible rank sum.
	}
	\label{fig2}
\end{figure}

\begin{figure}[htbp]
	\centering
		\includegraphics[width=0.8\linewidth]{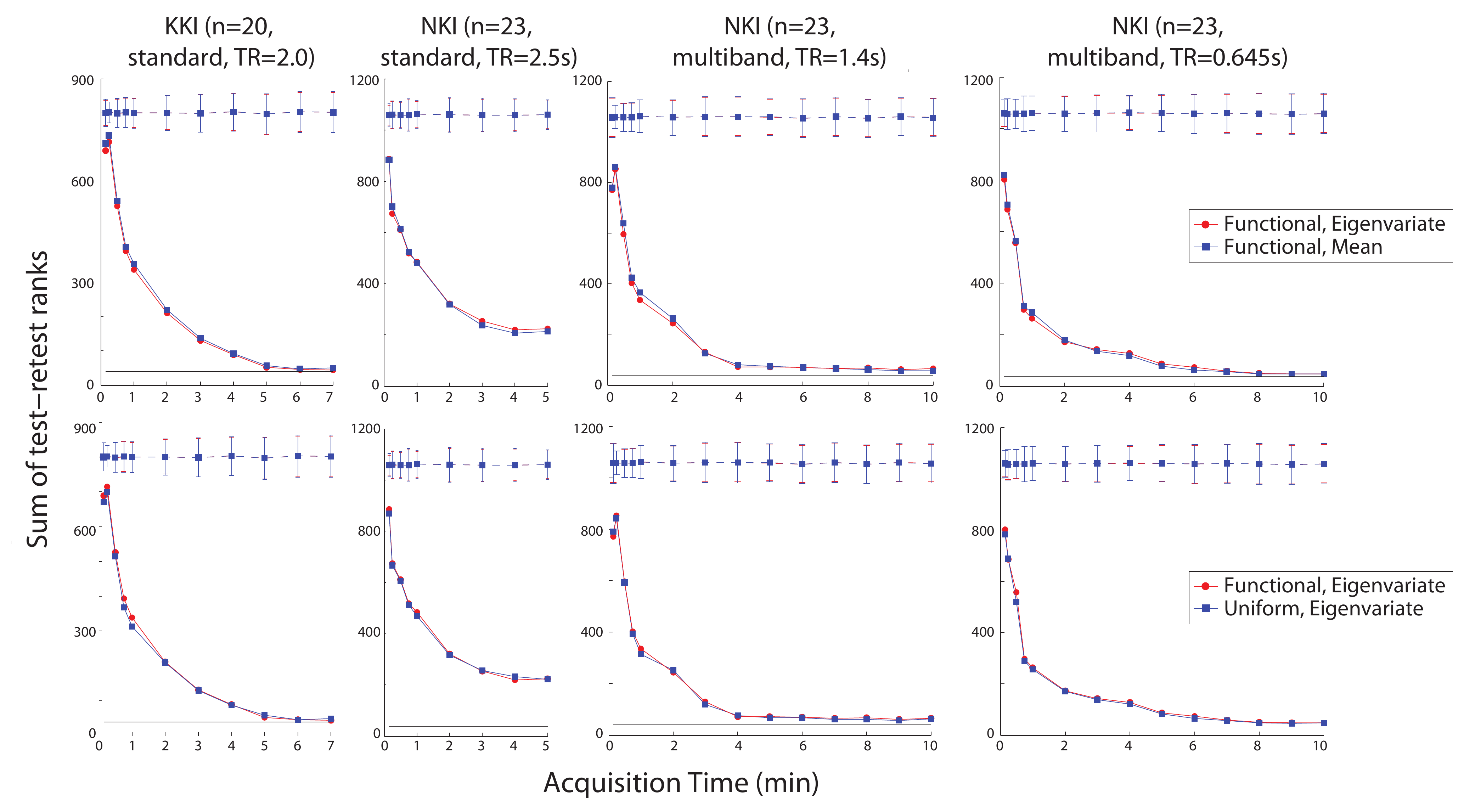}
	\caption{Varying methods of time-series extraction and parcellation generation for each test-retest data set.
		Top: Sum of test-retest scan pairs comparing method of timecourse-series extraction, plotted similar to Figure \ref{fig2}.
		Bottom: Sum of test-retest scan pairs comparing method of parcellation generation, plotted similar to Figure \ref{fig2}.
	}
	\label{fig3}
\end{figure}

We used minimization of this rank sum to define individual subject differentiability. For each analysis and acquisition parameter that was varied, several general trends were seen (Figures \ref{fig2} and \ref{fig3}). As expected, a longer acquisition time and a higher number of ROIs in the parcellation allowed greater ability to differentiate between individual subjects, with acquisition parameters (such as length of acquisition and TR) having greater effect. 

Acquisitions with shorter TR tended to allow greater individual subject differentiation, with the multiband acquisition data sets producing lower rank sums compared to standard acquisitions for the same acquisition duration. For the data sets analyzed, between 7-10 min of acquisition time was sufficient to minimize the resultant rank sum suggesting that acquisitions of a longer time period would produce only marginal results regarding individual subject characterization, for the number of subjects in these data sets (n=20  or 23, Figures \ref{fig2} and \ref{fig3}).

\begin{SCfigure}
	\centering
		\includegraphics[width=4in]{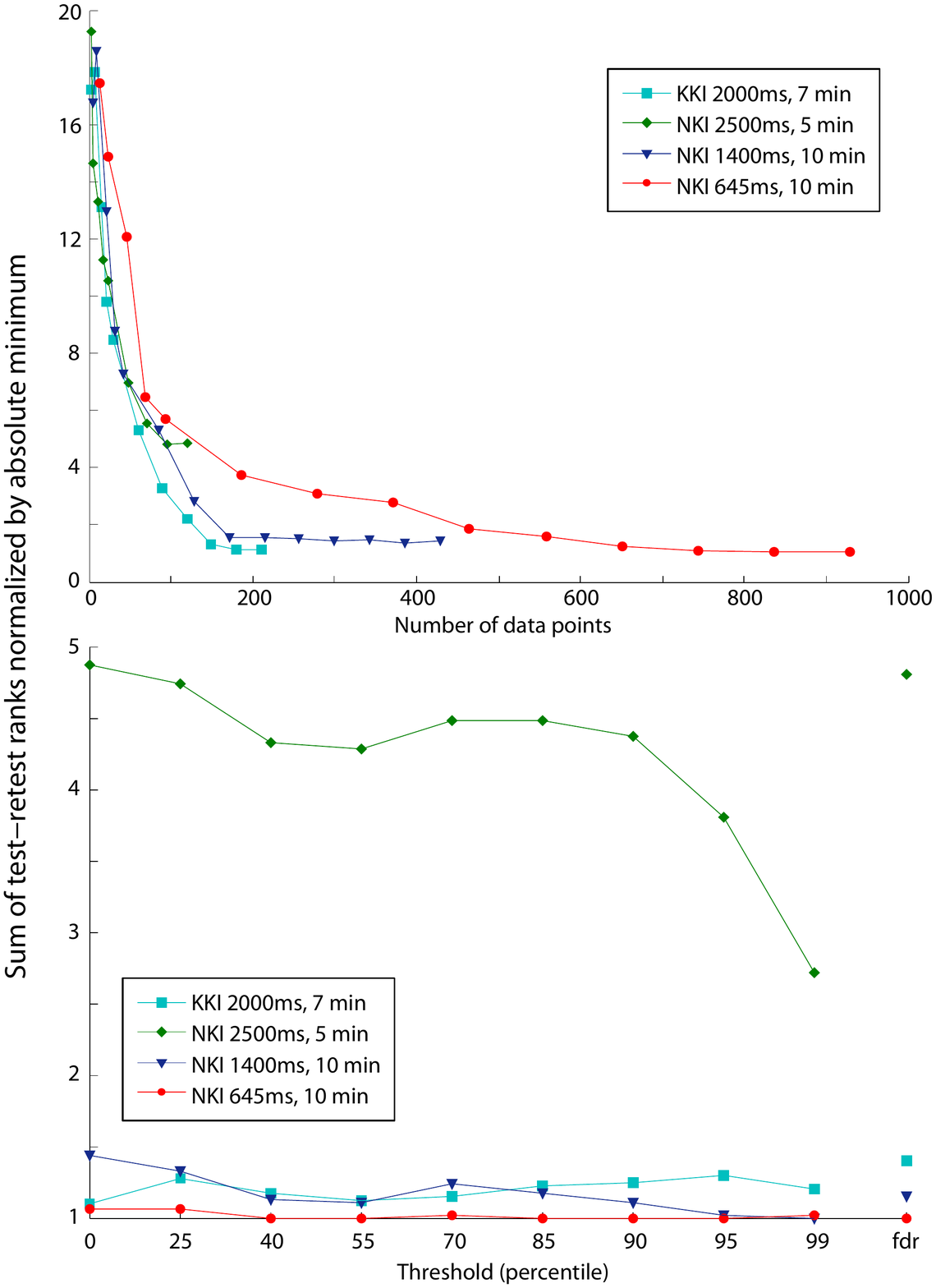}
	\caption{Factors affecting achieving the minimum rank sum.
		Top: Rank sums of test-retest pairs for each data set as a function of number of data points acquired.
		Bottom: Adjacency matrices were thresholded such that edges with correlation less than the indicated percentile were set to zero. Values presented are for the indicated data set processed using the functional parcellation of 1920 ROIs and using eigenvariate time-series extraction.}
	\label{fig4}
\end{SCfigure}

To determine whether the effect of TR on minimum rank sums was purely a function of increased sampling versus acquisition time, the data were organized by number of data points acquired, instead of by real time (Figure \ref{fig4}, top). These data seem to indicate that increased sampling frequency alone does not ensure lower rank sums. Instead, data sets with smaller TR tended to high rank sums when the number of data point were kept constant, suggesting an apparent trade-off of increased sampling frequency and longer acquisition times \cite{birn2013effect}.

As thresholding of the adjacency matrices is commonly utilized to reduce data dimensionality and eliminate likely noisy data, the minimum rank sum was calculated for each adjacency matrix after eliminating correlations below a percentile threshold (i.e. at a ``25\%'' threshold, all correlations below the 25th largest percentile of correlations were set to 0). With this analysis (Figure \ref{fig4}, bottom), the data set with the least sampling (NKI, standard acquisition with TR=2500 ms, 5 min acquisition) showed a steady improvement of subject differentiation with increasing thresholding. However, for data sets that otherwise achieved or were close to the minimum rank sum, there was no such effect. 

\subsection{Localization of Inter-Individual Differences}

To explore the question of the acquisition time necessary to differentiate individual subjects by their non thresholded similarity matrices, an unsupervised genetic algorithm was utilized to sort scans into their test-retest pairs using the rank sums as the optimization metric, without including any information other than the similarity matrices. Indeed, only 3-5 min of acquisition time was necessary to perfectly sort up to 23 subjects without any ad hoc subject labeling (Tables \ref{tab1} and \ref{tab2}). Choice of parcellation type produced only marginal differences with respect to time necessary for perfect pairing of subjects, as did number of ROIs in the parcellation beyond $\sim$1000.

\begin{table}
\caption{A genetic algorithm perfectly sorts true test-retest pairs using minimal acquisition time by minimizing the rank sum metric, with minimal variation with choice of parcellation. Elements of the table show the minimum time to perfectly sort scans from the indicated datasets and parcellation schemes.} \label{tab1}
\centering
\begin{tabular}{|c|c|c|c|c|}
	\hline
		Parcellation & functional & & uniform &  \\ \hline 
		Dataset & 1000 ROIs & 1920 ROIs & 1024 ROIs & 2048 ROIs \\ \hline 
		KKI (TR=2.0s) & 3 & 3 & 3 & 3 \\
		NKI (TR=2.5s) & 4 & 5 & 4 & 5 \\
		NKI (TR=1.4s) & 3 & 3 & 4 & 3 \\
		NKI (TR=0.645s) & 5 & 5 & 5 & 4 \\ \hline
\end{tabular}
\end{table}

\begin{table}
\caption{	A genetic algorithm perfectly sorts true test-retest pairs using minimal acquisition time by minimizing the rank sum metric, with minimally longer times needed with higher numbers of subjects.
	Minimum time for an unsupervised genetic algorithm to use test-retest rank sum minimization to perfectly sort scans from the indicated datasets into individual pairs, by number of subjects in the group (N). When varying N below the maximum of the data set, the set of subjects was randomly chosen from the total data set 20 times; presented are median values.
} \label{tab2}
\centering
\begin{tabular}{|c|c|c|c|c|}
	\hline
		Dataset & N=10 & N=15 & N=20 & N=23 \\ \hline 
		KKI (TR=2.0s) & 3 & 3 & 3 & N/A \\
		NKI (TR=2.5s) & 4 & 4 & 5 & 5 \\
		NKI (TR=1.4s) & 2 & 2 & 3 & 3 \\
		NKI (TR=0.645s) & 2 & 2 & 4.5 & 5 \\ \hline
\end{tabular}
\end{table}

\begin{figure}[htbp]
	\centering
		\includegraphics[width=1.0\linewidth]{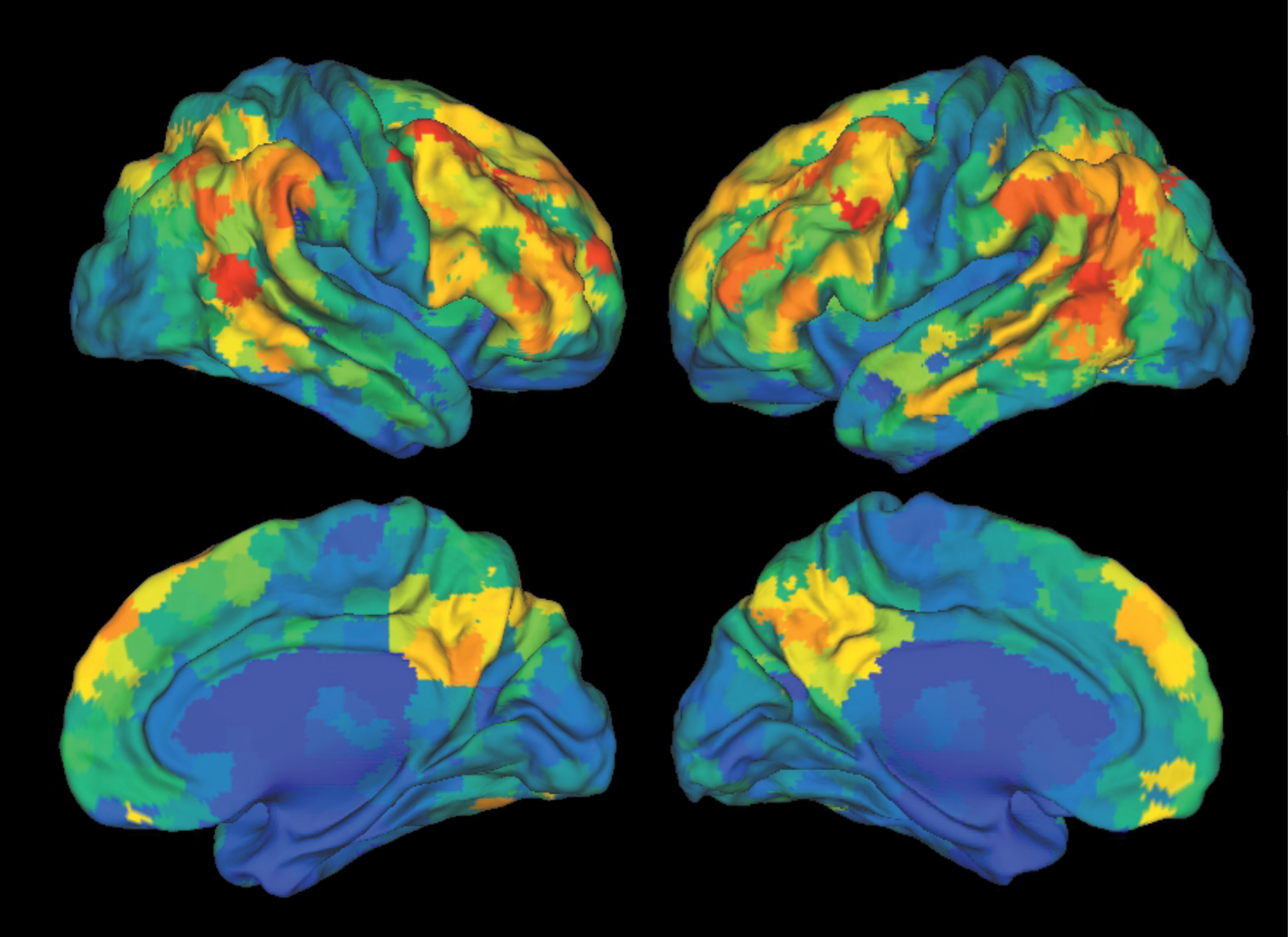}
	\caption{Map of brain regions that define unique identifiers of individuals
		Brain regions color coded by greatest number of
                connections with the lowest ($<$5th percentile) test-retest rank sum. Warmer colors code to higher individual differentiation (higher number of connections that have low test-retest rank sum).}
	\label{fig5}
\end{figure}

To further determine the brain regions and connections that contribute to this individual subject differentiation, for each dataset the rank sum metric was calculated for differences of each element of the adjacency matrix (using the functional parcellation, with the maximal number of ROIs, and eigenvariate time-series extraction). The brain regions containing the highest proportion of matrix elements with the lowest rank sum ($<$5th percentile) were determined (Figure \ref{fig5}). These regions were found to lie in secondary, association cortices including the parietal and prefrontal cortices and not in primary motor or sensory cortices. 

\section{Discussion}

Using the rank sum metric we describe (Figure \ref{fig1})---a nonparametric test of determining ability of a specific acquisition/analysis parameter set to differentiate individual subjects---we see several general trends. Perhaps unsurprisingly, more data allows for generally better individual differentiation (Figures \ref{fig2} and \ref{fig3}). For instance, longer acquisition times yield lower rank sums as do sequences with smaller TR. This effect seemed largely due to increased number of samples, although TR seemed to somewhat modulate these results (Figure \ref{fig4}, top). From our results,  it was not clear whether multiband acquisition alone had any advantages over standard acquisitions aside from the lower TR. 
Higher numbers of ROIs, with correspondingly lower size of individual ROIs, allowed for somewhat greater individual differentiation (Figure \ref{fig2}). This may be explained by less averaging together of dissimilar regions. 
Factors that had at most minor effect on individual subject differentiation were (i) choice of parcellation strategy, and (ii) time-series extraction method (Figure \ref{fig3}). 

For relatively undersampled datasets, thresholding the adjacency matrices (likely removing predominantly noise contributing elements) allowed for greater individual subject differentiation (Figure \ref{fig4}, bottom). However, this effect was not consistent with data sets with lower TR, suggesting that increased sampling alone could optimize the attainable inter-individual differentiation.
It was surprising that apparently limited amounts of data could allow for robust individual subject differentiation (Tables \ref{tab1} and \ref{tab2}). With only 2 min of BOLD acquisition time, using the multiband data sets, an unsupervised  algorithm could reliably and perfectly sort into the appropriate test-retest pairs up to 15 subjects---a typical value per group for many fMRI experiments. This sorting was able to occur even though the algorithm had no labels for which scan corresponded to which patient or even whether a particular scan was the initial test or the second, retest scan. The amount of acquisition time necessary for reliable, perfect, fully unsupervised sorting of scans increased for increasing numbers of subjects, up to only 5 min necessary for the full n=23 data sets. These results underscore that indeed rs-fMRI data alone contain sufficient information to robustly differentiate individual subjects and allow for analysis of the factors that contribute to individual subject uniqueness.

From the analysis of the brain regions and connections that most contribute to individual subject differentiation several interesting features were identified (Figure \ref{fig5}). First, primary sensory cortices and deep grey matter structures do not appear to contribute especially to individual subject differentiation. This is explained by the fact that such regions likely have an invariant functional anatomy and connectivity from person to person and therefore would not enable differentiation between individuals. In contrast, the regions that appear to contribute most to individual subject differentiation are found in association and secondary cortices in the prefrontal cortex, the precuneus and parietotemporal cortices. These latter cortical regions are thought to have undergone evolutionarily recent cortical expansion, supporting their putative role in higher cognitive processes \cite{BucknerRL2013}. Indeed, it is most likely that unique features of individuality would lie within these latter regions that are thought to contribute to higher-level association and conceptualization, and therefore more likely to be dependent on an individual's personality.

Of note, these latter regions are thought to comprise much of the default mode (DMN), attention (ATT) and executive control (EC) networks \cite{PowerJD2011}. These networks have been implicated in a heterogeneous array of interesting effects in the rs-fMRI literature. Our findings further suggest that these networks are the highest signal regions for determining the pertinent functional connectivity for an individual subject. As these regions display the greatest inter-subject variability, our findings warrant caution in interpretation of results that may average together functional connectivity statistics for these networks across a group of varied individuals.

It is unclear how much the step of normalization to an anatomic standard (completed as part of standard preprocessing) may affect these results and inform our determination of inter-individual differences. Currently such anatomic warping is a standard practice in processing both task-based and resting-state fMRI \cite{VanDijkKR2010}. It is possible that such warping may impart a signal in the derived functional connectivity that allows for individual subject differentiation based more on anatomy as opposed to fluctuations in neurovascular coupling. However, given that the typical native resolution of the BOLD imaging ($\sim$3 mm isotropic) is $\sim$25 fold less than the typical native resolution of the anatomic T1 acquisitions ($\sim$1 mm isotropic), we find it unlikely that the BOLD acquisition would have spatial resolution to distinguish these normal subjects based significantly on anatomy. There are no known methods for exactly comparing functional connectivity graphs of unwarped brains in such a manner as we have completed here. Indeed, this graph isomorphism problem is possibly NP complete for an exact solution \cite{Garey1979}. Certainly, future work will seek to complete a similar analysis, but for BOLD data that is not spatially warped to an anatomic standard.

Further avenues for research will be to further refine our map of brain regions and connections that are most stable across individuals versus within individuals. Additionally, it would be of interest to define a typical time length for this stability of individual functional connectivity patterns---in that certain regions and connections may vary across minutes, days, months or years depending on the changing cognitive state of the subject.

\section{Conclusion}

In this study, we have developed a non-parametric measure to evaluate the degree to which a given acquisition and analysis scheme can differentiate individual subjects. Using this metric, we see that there is a relative tradeoff of increased temporal sampling through either lower TR or high acquisition times. We further find that only 4-5 min of acquisition time is necessary to perfectly differentiate individual subjects using the described, standard methods. We find that brain regions that most contribute to this individual subject characterization lie in regions of higher cognitive processing. These results have application in study design for the question of analyzing individual subject level determinants of behavior and in the clinical evaluation of individual patients.

\bibliographystyle{elsarticle-num.bst}
\bibliography{Airan2014}

\section*{Acknowledgements}

This work is graciously supported by the Defense Advanced Research
Projects Agency (DARPA) SIMPLEX program through SPAWAR contract
N66001-15-C-4041 and DARPA GRAPHS N66001-14-1-4028 and NIH grant P41 EB015909.


\end{document}